\documentclass[11pt]{article}

\usepackage{epsfig}

\newcommand{\teflon}{Teflon$^{R}$ }

\begin{document}

\title{
Radioactive $^7$Be targets for measurements of the cross section of the $^7$Be(p,$\gamma$)$^8$B reaction
}
\author{ M. Hussonois, L. Brillard, C. Le Naour \\
IPN, IN2P3-CNRS et Universit\'{e} Paris-Sud, 91406 Orsay, France}


\maketitle

\section{Introduction}

The interpretation of the most recent solar neutrinos experiments requires a good knowledge of the 
cross section of the reaction $^7$Be(p,$\gamma$)$^8$B at very small energy (E$_{cm}$=18 keV).
We have recently measured\cite{HAM98}
this cross section for E$_{cm}$=0.35-1.4 MeV and 
for E$_{cm}$=0.112-0.190 MeV. We report here on the description of the preparation of the radioactive 
targets of $^7$Be used in these experiments.

\section{$^7$Be build up}
$^7$Be was produced via the $^7$Li(p,n)$^7$Be reaction at the 4 MV Van de Graaff accelerator in Bordeaux.
The enriched (99\%) metallic $^7$Li target was 7 mg/cm$^2$ thick and was deposited on a water-cooled backing.
The irradiation was carried on for about 30 days, with an average beam intensity of about 30$\mu$A at a proton energy 
of 3.2 MeV. At the end of the irradiation, the $^7$Li target was kept under vacuum until chemical processing begins.

\section{Chemical procedure.}

\subsection{Materials.}

The contamination by macro-amounts of alkali earth elements (such as Ca, Ba),
homologue of Be and difficult to separate, was avoided by use of vessels in
\teflon or in quartz, perfectly cleaned by boiling with ultra-pure nitric
acid and water.

All the concentrated acids and water, already of high purity, were distilled,
just before use, in \teflon apparatus.  Diluted acids were prepared in
quartz vials. The three used columns were made with \teflon tubes (2.4 mm inner
diameter) and \teflon connectors.  The different acids were flowed
through these columns with a peristaltic pump by means of tubings in \teflon
and/or Tygon. 

The first column was filled with the di-2-ethyl hexyl
phosphoric acid (HDEliP) sorbed on \teflon powder (Voltalef, particle
diameter less than 50$\mu$m).  After washing with boiled distilled water (to
eliminate air bubbles and excess of HDEHP), this column was cleaned from
mono, di and trivalent elements by washing with 10 ml of 6M HN0$_3$.  The
minimum of water was used to wash till  neutrality, and finally the
column was conditioned with 2 ml of 5. 10$^{-3}$ HNO$_3$.

The second column was filled with the cation exchange resin AG 50 X 8, 
200 - 400 mesh previously treated 
to eliminate Fe impurities.  Nevertheless, the column was successively
washed with 6M HN0$_3$, water and 0.1 M HN0$_3$. The third one is a small column
filled with about 50 mg of macro-porous anion exchange resin, AGMP1, which
gives very few products of degradation with concentrated acid.  This
column was previously washed with water and 8M HCl. 

The different steps of the chemical separation were monitored by measurements 
of all fractions by detecting the 478 keV $\gamma$-ray of $^7$Be by $\gamma$-spectroscopy chain.

\section{Target dissolution.}

The target, lithium deposit on its copper backing, conserved under vacuum
till the beginning of the chemistry, is rapidly placed in the dissolution
cell.  This system, practically closed, was developed to avoid
contamination.  It is composed, at the bottom, of a \teflon ring, of inner
diameter adjusted to the size of the lithium deposit, which prevent leakage
and contact with copper and glass of the alcohol-solvent.  The upper part is
a glass tube, closed up by a \teflon flask with a small hole (diameter= 3mm) 
to introduce and pick up the solvent.  This ensemble is pressed between two
copper flasks. 

The dissolution starts immediately with the introduction of a
small volume (1 ml) of pure methanol.  The thickness of the \teflon ring is
enough to prevent escape of small droplets generated by bubbling.  When
this bubbling diminishes, the alcohol is picked up and replaced by a new
fresh portion.  This cycle is repeated till complete dissolution of the
lithium, taking care of a possible attack of the copper support by the
lithium hydroxide. All the alcoholic fractions are collected in a quartz
beaker, in presence of 0.5ml of distilled nitric acid, to prevent hydrolysis
of beryllium in presence of LiOH.  After evaporation by infra-red
heating of the alcohol, all traces of its, which can wash out HDEHP from the
column, are eliminated by three successive evaporations to dryness of
concentrated HN03.  Finally the residue is dissolved in 35 ml of 5.10$^{-3}$M
HNO$_3$, to have a low lithium concentration.

\section{Chemical separations.}

The first step is the elimination of the bulk of lithium, based on the
difference of extrabilities of Li and Be by HDEHP from 5.10$^{-3}$ M HN03
solution.  The percolation of the previous nitric solution, at a low flow
rate of 1 ml per 5 minutes, fixes the $^7$Be on the column when Li passes through.
The effluent is collected by 5 ml fractions, the low activity of each fraction being
controlled. After percolation of the initial solution, the beaker is  decontaminated 
three times with 2 ml of 5. 10$^{-3}$ M HN03 and finally the HDEHP
column is washed by two times 5 ml of pure 5.10$^{-3}$ M HN03. Better than the
slow elution with 4M HN03, due certainly to Be hydrolysis, the elution with
1 M hydrofluorhydric acid permits to recover more than 90\% of the $^{7}$Be
activity in the first milliliter, 95\% in two milliliters and 98\% in four
milliliters, important to minimize the introduction of impurities in the $^{7}$Be
fraction. The second purification on a cationic exchanger resin is necessary
to eliminate traces of HDEHP washed out with HF.  So the different HF
fractions are successively evaporated in a \teflon cones.  To eliminate HF
traces, three times 0.5 ml of concentrated HN03 are evaporated to dryness.

The $^7$Be is then dissolved in the cones by five successive additions of 0.4 ml of
0.4 M HN03 and directly picked from the cones to the
$^7$Be desorption which is realized by 3 ml of 2M HN03 collected directly in a new \teflon cone.

The evaporation to dryness, the pick-up of the $^7$Be 
by 3 times 0. 1 ml of 8M HCl and the flow of this solution through
the small AGMP1  column ensure complete elimination of iron impurities
 observed in the previous tries despite all  precautions. 

The yield of all these separation processes is greater than 90\%

\boldmath

\section{ $^7$Be target.}

\unboldmath

The preparation of an uniform thin deposit of a high yield of $^7$Be 
on a thick platinum disk($\phi$ = 20mm, e = 0.2 mm) was developed by molecular plating
of $^7$Be nitrate which was dissolved in 2-methyl- 1 propanol, 99.5\% HPLC grade
(Sigma-Aldrich Company). 

So, after addition of 0.4 $\mu$g of $^9$Be carrier (from
beryllium atomic absorption standard solution, Aldrich Chemical Company,
Inc.), the 8M HCl solution of $^7$Be is evaporated to dryness in a new Teflon
cone.  The chloride is converted into nitrate by two successive
evaporations of 0.2ml of distilled nitric acid.  Finally the Be is picked
up by three times 6 micro-liters of 10$^{-1}$M HN03 mixed with
1.2 ml of 2-methyl-1 propanol.

 This solution is transfered into the
electrolytic cell, presented in fig. 1. A \teflon ring defines the volume
of the cell and the diameter of the deposit ($\phi$ = 8 mm).  The two parallel
platinum electrodes are strongly pressed on this ring, to prevent 
leakage, by two metallic pieces protected by Afcodur$^{R}$ insulators. After 45
minutes of electrolysis, with a 740V applied voltage and a 5 mA current,
more than 95\% of the $^7$Be activity is deposited.  After settling of
the solution, the platinum disk is recovered and flamed to red color, to
obtain  beryllium oxide. The total yield of the operation, starting from the 
beginning is about 85\%.

Two $^7$Be targets were prepared in this manner: the first one had an activity of 26.9 mCi and the second 
one of 131.7 mCi.In both cases the deposit surface was about 0.5 cm$^2$.

\begin{center}

\mbox{\epsfig{file=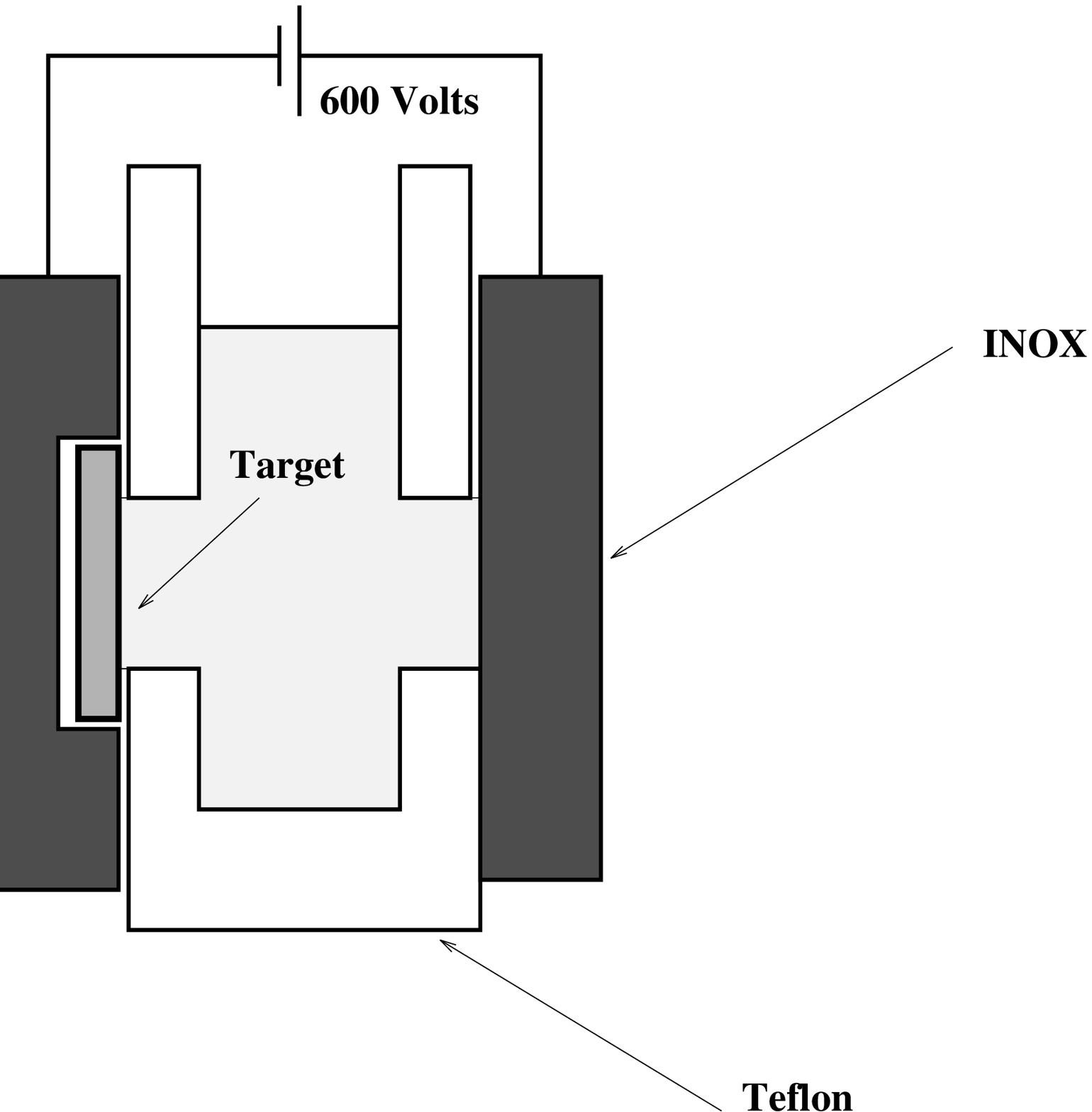,width=10.cm}}

{\small \textsl{
Figure 1
Sketch of the device for the electrolysis\ }}

\end{center}

\end{document}